\documentclass[twocolumn,superscriptaddress,showpacs,prl]{revtex4}

\usepackage[dvips]{graphicx,color}
\usepackage{delarray}
\usepackage{amsmath}
\usepackage{bm}

\newcommand{\braket}[2]{\langle {#1} | {#2} \rangle}
\newcommand{\ket}[1]{| { #1} \rangle}
\newcommand{\bra}[1]{ \langle {#1}  |}

\begin{document}
\title{Optimal entanglement manipulation via coherent-state transmission}

\author{Koji Azuma}
\email{azuma.koji@lab.ntt.co.jp}
\affiliation{NTT Basic Research Laboratories, 3-1 Morinosato Wakamiya, Atsugi, Kanagawa 243-0198, Japan}

\author{Go Kato}
\email{kato.go@lab.ntt.co.jp}
\affiliation{NTT Communication Science Laboratories, 3-1 Morinosato Wakamiya, Atsugi, Kanagawa 243-0198, Japan}

\date{\today}

\begin{abstract}
We derive an optimal bound for arbitrary entanglement manipulation based on the transmission of a pulse in coherent states over a lossy channel followed by local operations and unlimited classical communication (LOCC).
This stands on a theorem to reduce LOCC via a local unital qubit channel to local filtering.
We also present an optimal protocol based on beam splitters and a quantum nondemolition (QND) measurement on photons. 
Even if we replace the QND measurement with photon detectors, the protocol outperforms known entanglement generation schemes.
\end{abstract}
\maketitle


Entanglement is now well known as an essential resource for quantum communication \cite{B93} 
despite it being found in an attempt to point out a paradoxical nature of quantum mechanics \cite{EPR}.
In fact, it is known that any quantum communication [including quantum key distribution (in the sense of Ref.~\cite{BBM92})] can never be accomplished by 
distant parties who are not capable of sharing entangled pairs.
This implies the importance of evaluating the potential to share entanglement through a given communication channel, which determines its value as a quantum channel.
If we look at practical quantum communication such as fiber-based quantum key distribution, free-space quantum communication, entanglement generation in quantum repeaters, quantum communication via superconducting transmission lines, and a quantum memory for bosons (transmission in time), we become aware that all the protocols rely on a lossy bosonic channel.
Thus, quantum communication based on this channel is practically the most important class (cf.~\cite{WPG07}).

One of the most fundamental protocols in this class is
the family of coherent-state-based protocols represented by Bennett 1992 quantum key distribution \cite{B92} and entanglement generation protocols in quantum repeaters \cite{C06,L06,L08,M08,A09,ATKI10}.
These protocols are based on the transmission of a pulse in coherent states over a lossy channel, and they are dominated by the following paradigm: (i) A sender prepares an entangled state composed of computational basis states of a qubit $A$ and coherent states of a pulse $a$. (ii) The sender then sends the pulse $a$ to the mode $b$ at the receiver's cite through a lossy channel. (iii) Then, the sender and the receiver manipulate the shared system $Ab$ through their local operations and unlimited two-way classical communication (LOCC)
in order to convert the initial entangled state to a more entangled state by tolerating failure.
Hence, the potential of the coherent-state-based protocols is determined by optimizing the LOCC manipulation for a single entangled pair $Ab$.
This kind of ``entanglement manipulation'' is completely understood  for a pure-state input $Ab$ \cite{JP99}. But, the analysis for a mixed-state input $Ab$ as considered here has remained a long-standing open question \cite{JP99} despite its significance.
In addition, the LOCC manipulation is beyond the paradigms in Refs.~\cite{L08,A09,A10}.
Therefore, grasping the potential of such coherent-state-based protocols must be a key step in the practical and theoretical evolution of quantum communication.

In this paper, we present a theoretical limit of the performance of arbitrary coherent-state-based protocols, as well as a simple protocol that achieves the limit.
This is based on a general proposition to reduce LOCC manipulation via a local unital qubit channel to local filtering. 
The derived limit is represented in terms of the total success probability and an average entanglement monotone \cite{V00} of the generated entangled states, and it is determined only by the transmittance of the channel.
The bound is shown to be accomplished by a simple protocol composed only of 
beam splitters and a quantum nondemolition (QND) measurement \cite{BVT80} on photons.
If we substitute photon-number-resolving detectors for the QND measurement, the protocol can entangle distant qubits with near-optimal performance, which is shown to outperform known protocols \cite{C06,L06,L08,M08,A09}.
Hence, these protocols play the role of an efficient entanglement supplier for various quantum communication schemes.

{\it Coherent-state-based protocols.}---We start by defining the protocols considered here:
(A-i) A sender called Alice prepares a qubit $A$ and a pulse $a$ in her desired state in the form of $ \sum_{j=0,1} e^{i \Theta_j} \sqrt{q_j} \ket{j}_A \ket{\alpha_j}_a$ for a computational basis $\{ \ket{j}_A \}_{j=0,1}$, coherent states $\{\ket{\alpha_j}_a\}_{j=0,1}$, real parameters $\Theta_j$, and $q_j \ge 0$ with $\sum_{j=0,1} q_j=1$;
(A-ii) Alice sends the pulse $a$ to a receiver called Bob, through a lossy channel described by an isometry
$\ket{\alpha}_a \to \ket{ \sqrt{T} \alpha}_b \ket{\sqrt{1-T} \alpha}_{e}$,
where $T$ is the transmittance, $b$ is a mode at Bob's place, and $e$ is the environment;
(A-iii) Then, Alice and Bob manipulate the system $Ab$ through LOCC to obtain an entangled state $\hat{\tau}_k^{A'B}$ between Alice's system $A'$ and Bob's system $B$, and declare whether they obtain a success outcome $k$ occurring with a probability $p_k$ or a failure outcome. Note that the output systems $A' B$ are not limited to qubits \cite{cont}. 
In what follows, the set of all the success events $k$ is denoted by ${\cal S}$.

As a measure of the performance of the protocols, we take the total success probability, i.e.,
$
P_s=\sum_{k \in {\cal S}} p_k.
$
We also need to choose an entanglement measure for estimating the value of the obtained entangled states $\{\hat{\tau}_k^{A'B}\}_{k \in \cal S}$. Since the output system $A'B$ has no restrictions in contrast to those described in Refs.~\cite{L08,A09,A10}, the singlet fraction may be unsuitable.
Thus, here we take an entanglement monotone $E$ \cite{V00} that is a convex monotonically nondecreasing function of the concurrence $C$ \cite{W98} at least for qubits (cf.~\cite{examples}).
Based on this $E$, as another measure of the protocols, 
we adopt the average $\bar{E}$ of the obtained entangled states $\{\hat{\tau}_k^{A'B}\}_{k\in {\cal S}}$, namely
$
\bar{E}= [\sum_{k\in {\cal S}} p_k E (\hat{\tau}_k^{A'B})]/P_s
$.

We also allow Alice and Bob to switch among two or more protocols probabilistically. This corresponds \cite{A10} to taking the convex hull of achievable points $(P_s, P_s \bar{E})$.

{\it Virtual protocol.}---For an actual protocol, we define the virtual protocol \cite{A09} that works in the same way as the actual protocol but simplifies the analysis significantly.
Steps (A-i) and (A-ii) indicate that, when the pulse arrives at Bob's site, the state of the total system $A b e$ is written in the form
$
 \ket{\psi}_{A b e} = \sum_{j=0,1} \sqrt{q_j} \ket{j}_A \ket{u_j}_{b}
  \ket{v_j}_e 
\label{eq:abe}
$
for states $\{\ket{u_j} \}_{j=0,1}$ and $\{\ket{v_j} \}_{j=0,1}$ with
$
|\braket{u_1}{u_0}|^{1-T} = |\braket{v_1}{v_0}|^T>0.
 \label{tra}
$
Thus, for a state 
$
\ket{\psi'}_{A b}:= \sum_{j=0,1}  \sqrt{q_j} e^{i (-1)^j  \xi } \ket{j}_A \ket{u_j}_{b} 
$
with $2 \xi:=\arg [\braket{v_1}{v_0}]$ and for a phase-flip channel
$
\Lambda^A_{u} (\hat{\rho}):=f_{u} \hat{\rho} + (1-f_{u}) \hat{Z}^A \hat{\rho}
\hat{Z}^A \label{Lambda}
$
with $\hat{Z}^A := \ket{0} \bra{0}_{A} -\ket{1} \bra{1}_{A}$ and
$
f_{u}:=(1+u^{\frac{1-T}{T}})/2,
$
we have
$
{\rm Tr}_{e} [ \ket{\psi}\bra{\psi}_{A b e} ] = \Lambda^A_{|\braket{u_1}{u_0}|}(\ket{\psi'} \bra{\psi'}_{A b}). \label{eq:red}
$
Hence,
we can consider
any protocol to have the following sequence:
(V-i) System $A b$ is prepared in $\ket{\psi'}_{A b}$;
(V-ii) $\Lambda^A_{|\braket{u_1}{u_0}|}$ is applied on qubit $A$;
(V-iii) Alice and Bob perform an LOCC, which provides $\hat{\tau}_k^{A'B}$.
We call this sequence ``the virtual protocol.''

We introduce a proposition that enables us to derive an optimal bound in more general settings (cf.~\cite{converse}).

{\it Proposition.}---Let $(P_s, \bar{E})$ be the performance of an LOCC protocol starting with qubits $AB$ in state $ {\cal E}^A(\ket{\varphi}\bra{\varphi}_{AB})$, where ${\cal E}^A$ is a random local unitary channel \cite{unital} defined by ${\cal E}^A(\hat{\rho}^{AB}):=\sum_l q_l \hat{U}_l^{A} \hat{\rho}^{AB} (\hat{U}_l^A)^\dag$.
Then, there is a protocol that is not less efficient than $(P_s,\bar{E})$ but that is based only on Bob's measurement.
In addition, for Schmidt coefficients $\lambda_0$ and $\lambda_1(\le \lambda_0)$ of $\ket{\varphi}_{AB}$, 
the achievable region of $(P_s, P_s \bar{E})$ is described by the convex hull of $\{(P_s, P_s \bar{E})\;|\; 0 \le P_s \le 1,0 \le \bar{E} \le E(C^{\rm max}(P_s)) \}$ with $C^{\rm max}(P_s):=(2 \sqrt{\lambda_0 \lambda_1})^{-1} C({\cal E}^A(\ket{\varphi}\bra{\varphi}_{AB})) $ for $P_s < 2 \lambda_1$ and $C^{\rm max}(P_s):= P_s^{-1}\sqrt{(P_s-\lambda_1)/ \lambda_0} C({\cal E}^A(\ket{\varphi}\bra{\varphi}_{AB}))$ for $P_s \ge 2 \lambda_1 $.

{\it Proof.}
Let Kraus operators $\{\hat{M}_k^A \otimes \hat{N}_k^B \}_{k \in {\cal S}}$ be Alice and Bob's successful measurement in step (V-iii).
Without loss of generality, the input spaces of $\hat{M}_k^A$ and $\hat{N}_k^B$ can be assumed to be qubit spaces.
If Alice and Bob can achieve the measurement $\{ \hat{M}_k^A \otimes \hat{N}_k^B \}_{k \in {\cal S}}$, they can always, in principle, obtain a state 
$\hat{\tau}_k^{AB} := (\hat{M}_k^A \otimes \hat{N}_k^B) {\cal E}^A(\ket{\varphi} \bra{\varphi}_{AB}) (\hat{M}_k^A \otimes \hat{N}_k^B)^\dag /p_k$.
From the the convexity of the entanglement monotone $E$ \cite{V00}, 
the performance of this protocol is not less than protocols where, for a set ${\cal S}' \subset {\cal S}$, they provide a mixture of the states $(\sum_{k \in {\cal S}'} p_k \hat{\tau}_k^{AB})/(\sum_{k \in {\cal S}'} p_k)$ instead of states $\{\hat{\tau}_k^{AB}\}_{k \in {\cal S}'}$.
Thus, we can assume that Alice and Bob return the state $\hat{\tau}_k^{AB}$ with probability $p_k$.
Note that the range of $\hat{\tau}_k^{AB}$ can be assumed to be qubit spaces.

From Proposition 1 in Ref.~\cite{LP01}, for any $\hat{U}^A_l$, there exist unitary operators $\{\hat{V}^A_{k|l} \}_k$ and Kraus operators $\{ \hat{O}_{k|l}^B\}_{k}$ that satisfy
$(\hat{M}_k^A \hat{U}^A_l \otimes \hat{N}_k^B ) \ket{\varphi}_{AB} = (\hat{V}_{k|l}^A \hat{U}^A_l \otimes \hat{O}_{k|l}^B ) \ket{\varphi}_{AB}$ with $d_k:={\rm det}[(\hat{M}_k^A)^\dag \hat{M}_k^A] {\rm det}[(\hat{N}_k^B)^\dag \hat{N}_k^B] = {\rm det}[(\hat{O}_{k|l}^B)^\dag \hat{O}_{k|l}^B]$.
On the other hand, using the formula \cite{W98}, we can show that the concurrence $C$ for the state $\hat{\tau}_k^{AB}$ is described by $p_k C(\hat{\tau}_k^{AB})=\sqrt{d_k} C({\cal E}^A(\ket{\varphi}\bra{\varphi}_{AB}))$.
Thus, if Bob performs $\{ \hat{O}_{k|l}^B\}_{k}$,
he obtains a state $\hat{\tau}_{k|l}^{AB}:= \hat{O}_{k|l}^B {\cal E}^A(\ket{\varphi}\bra{\varphi}_{AB}) (\hat{O}_{k|l}^B)^\dag /p_{k|l}$ with probability $p_{k|l}:=\bra{\varphi} (\hat{O}_{k|l}^B)^\dag \hat{O}_{k|l}^B \ket{\varphi}$ and concurrence $C(\hat{\tau}_{k|l}^{AB})=\sqrt{d_k} C({\cal E}^A(\ket{\varphi}\bra{\varphi}_{AB}))/p_{k|l}= p_k C(\hat{\tau}_{k}^{AB}) / p_{k|l}$.
Since $\sum_l q_l p_{k|l}=p_k$ holds and $\sum_l q_l p_{k|l} E(\tau^{AB}_{k|l}) \ge  p_k E(\tau^{AB}_{k}) $ is implied by $\sum_l q_l p_{k|l} C(\hat{\tau}^{AB}_{k|l}) = p_k C(\hat{\tau}^{AB}_{k}) $ and the convexity of $E(C)$, 
the original LOCC protocol is concluded to be outperformed by a protocol 
that performs only Bob's measurement $\{ \hat{O}_{k|l}^B\}_{k}$ with probability $q_l$ and returns $k$ and $l$ as the outcome. 

Thus, we focus on a protocol that is based on Bob's measurement $\{\hat{O}_{k}^B \}_{k \in {\cal S}}$ and returns state $\hat{\rho}_k^{AB}:=\hat{O}_k^B {\cal E}^A(\ket{\varphi}\bra{\varphi}_{AB}) (\hat{O}_k^B)^\dag/p_k $ with probability $p_k$.
We note that there are Kraus operators $\hat{\Omega}^B$ and $\{ \hat{L}_k^B \}_{k \in  {\cal S}}$ satisfying $\hat{L}^B_k \hat{\Omega}^B=\hat{O}_k^B$.
In fact, if we define them as $\hat{\Omega}^B:= [\sum_{k \in {\cal S}} (\hat{O}^B_k)^\dag \hat{O}_k^B ]^{1/2} $ and $\hat{L}^B_k:= \hat{O}_k^B (\hat{\Omega}^B)^{-1}$, where $\hat{\Omega}^{-1}$ is the inverse of $\hat{\Omega}$ in its range,
the operators satisfy $(\hat{\Omega}^B)^\dag\hat{\Omega}^B \le \hat{1}^B$ and $\sum_{k \in {\cal S}} (\hat{L}^B_k)^\dag \hat{L}^B_k \le \hat{1}^B $ from $\sum_{k \in {\cal S}} (\hat{O}^B_k)^\dag \hat{O}^B_k \le \hat{1}^B $.
Hence, we can regard Bob's measurement $\{ \hat{O}_k^B\}_{k \in {\cal S}}$ as a sequential measurement of $\hat{\Omega}^B$ followed by $\{ \hat{L}_k^B \}_{k \in {\cal S}}$.
On the other hand, the entanglement monotone $E$ of the state
$\tau_s^{AB}:=\hat{\Omega}^B {\cal E}^A(\ket{\varphi}\bra{\varphi}_{AB}) \hat{\Omega}^B/P_s$ with $P_s = \sum_{k \in {\cal S}} p_k$ is not less than $[\sum_{k \in {\cal S}} p_k E(\rho_k^{AB})]/P_s$, because the entanglement monotone $E$ does not increase through a local operation on average \cite{V00}. 
Therefore, we can assume that Bob merely applies a filter $\hat{\Omega}^B$ to qubits $AB$.

Let us proceed to the optimization of $(P_s, E(\tau_s^{AB}))$ over the filter $\hat{\Omega}^B$. 
From the monotonicity of $E(C)$, our attention is concentrated on the maximization of $C(\tau_s^{AB})$ for a fixed $P_s$.
On the other hand, for the Schmidt decomposition of $\ket{\varphi}_{AB}=\sum_{j=0,1} \sqrt{\lambda_j }\ket{jj}_{AB}$,
we have $P_s=\bra{\varphi} (\hat{\Omega}^B)^\dag \hat{\Omega}^B \ket{\varphi} = \sum_{j=0,1} \lambda_j \bra{j} (\hat{\Omega}^B)^\dag \hat{\Omega}^B \ket{j} $ and $P_s C( \tau_s^{AB}) = ({\rm det} [(\hat{\Omega}^B)^\dag \hat{\Omega}^B] )^{1/2} C({\cal E}^A(\ket{\varphi}\bra{\varphi}_{AB}) ) \le  (\Pi_{j=0,1} \bra{j} (\hat{\Omega}^B)^\dag \hat{\Omega}^B \ket{j} )^{1/2} C({\cal E}^A(\ket{\varphi}\bra{\varphi}_{AB}) )$, where the equalities hold by choosing $\hat{\Omega}^B$ with $\bra{0} (\hat{\Omega}^B)^\dag \hat{\Omega}^B  \ket{1}=0$.
Combined with $(\hat{\Omega}^B)^\dag\hat{\Omega}^B \le \hat{1}^B$, this shows that $C^{\rm max}$ is the maximum of $ C( \tau_s^{AB})$. By considering the mixture of protocols, the overall statement becomes the proposition.

{\it Optimal bound.}---Let us apply the proposition to our problem.
Schmidt coefficients of $\ket{\psi'}_{Ab}$ are $\lambda_{\pm}:=[1 \pm \sqrt{1-x^2}]/2$,
and the concurrence of the input state is $C(\Lambda^A_{|\braket{u_1}{u_0}|}(\ket{\psi'}\bra{\psi'}_{Ab}))=  |\braket{u_1}{u_0}|^{\frac{1-T}{T}}  x$ from Ref.~\cite{Appendix}, where $x:=2 \sqrt{q_0 q_1 (1- |\braket{u_1 }{u_0}|^2)}$. Hence, $C^{\rm max} (P_s) =|\braket{u_1}{u_0}|^{\frac{1-T}{T}}$ for $P_s < 1-\sqrt{1-x^2}$ and 
$C^{\rm max} (P_s) =P_s^{-1} |\braket{u_1}{u_0}|^{\frac{1-T}{T}} x [1-2 (1-P_s) /(1+\sqrt{1-x^2})]^{1/2}$ for $P_s \ge 1-\sqrt{1-x^2}$.
Since $C^{\rm max} (P_s)$ is a monotonically nondecreasing function of $x$, the choice of $q_0=q_1=1/2$ gives the maximum value of $C^{\rm max} (P_s)$, which is further bounded by an achievable concurrence $C^{\rm opt}_{u^*}(P_s)$ with
\begin{equation}
C^{\rm opt}_u(P_s):= \frac{ u^{\frac{1-T}{T}}  \sqrt{(1-u)(2P_s +u-1)}}{P_s}
\label{eq:upper1}
\end{equation}
for
\begin{equation}
u^{*}:= \frac{1}{2} \left[(1-P_s) (2-T)
+\sqrt{4 P_s^2 (1-T)+(1-P_s)^2 T^2} \right] \label{eq:uopt}
\end{equation}
satisfying $1-P_s \le u^* \le 1$.
Therefore, the performance $(P_s,P_s \bar{E})$ of any protocol must be in the convex hull of $\{(P_s,P_s \bar{E}) \;|\; 0 \le P_s \le 1,0 \le \bar{E} \le E(C^{\rm opt}_{u^*}(P_s)) \}$.

\begin{figure}[t]
  \begin{center}
    \includegraphics[keepaspectratio=true,height=65mm]{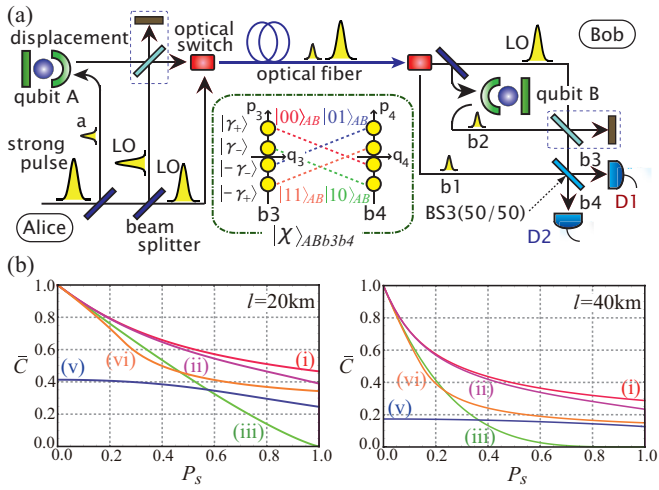}
  \end{center}
   \caption{
(a) Schematic of near-optimal protocol. If we replace the photon detectors D1 and D2 with the QND measurement to perform the projection $\hat{Q}_s^{b_3 b_4}$, we can reduce the protocol to the optimal one.
(b) Performance of various protocols: The average concurrence $\bar{C}$ as a function of the success probability $P_s$ when $T=e^{-l/l_0}$ with $l_0=25$ km ($\sim 0.17$ dB/km attenuation) 
and $\theta=0.01$, for (i) the optimal protocol, (ii) the near-optimal protocol, (iii) a photon-detector-based two-probe protocol \cite{A09} that achieves a tight bound \cite{A10} for single-error-type entanglement generation, (iv) a photon-detector-based single-probe protocol \cite{L08,M08}, and (v) a homodyne-detection-based single-probe protocol \cite{L06}.}
 \label{lo1}
\end{figure}

{\it Optimal protocol.}---We have shown that the achievable region of an arbitrary protocol is described by Eqs.~(\ref{eq:upper1}) and (\ref{eq:uopt}).
Here we present a specific protocol achieving the optimal bound $C^{\rm opt}_{u^*}(P_s)$ except for a trivial point $P_s=1$.
We allow Alice and Bob to use a realizable \cite{L06} interaction between an off-resonance laser pulse in a coherent state $\ket{\alpha}_a$ and a matter qubit $A$, which is described by a unitary operation 
$
\hat{U}_{\theta} \ket{j}_{A} \ket{\alpha}_a = \ket{j}_A \ket{\alpha e^{i (-1)^j \theta/2}}_a 
$
for $j=0,1$. $\theta$ depends on the strength of the interaction ($\theta \sim 0.01$ \cite{L06}).
Let us consider the following protocol [see Fig.~\ref{lo1} (a)]:
(1) Alice makes a probe pulse in a coherent state $\ket{\alpha/\sqrt{T}}_a$ ($\alpha \ge 0$) interact with her qubit $A$ in a state $(\sum_{j=0,1} e^{- i (-1)^j \zeta_{\alpha/\sqrt{T}} } \ket{j}_A)/\sqrt{2}$ with $\zeta_\alpha:= (1/2)\alpha^2 \sin \theta $ by $\hat{U}_\theta$,
and she applies a displacement operation $\hat{D}_{- (\alpha/\sqrt{T}) \cos (\theta/2) }$ to the pulse $a$;
(2) Alice sends the pulse to Bob through a lossy channel $a\to b_1$ (with transmittance $T$) together with the local oscillator (LO);
(3) On receiving the pulse $b_1$ and the LO, Bob generates a second probe pulse $b_2$ in a coherent state $\ket{\beta}_{b_2}$ with $\beta \ge \alpha$ from the LO, and he
makes the pulse $b_2$ interact with his qubit $B$ in state $(\sum_{j=0,1} e^{- i (-1)^j  \zeta_\beta  } \ket{j}_B)/\sqrt{2}$ by $\hat{U}_\theta$;
(4) Bob applies a displacement operation $\hat{D}_{-\beta \cos (\theta/2)}$ to the pulse $b_2$;
(5) Bob further applies a 50/50 beam splitter described by $\ket{\alpha_1}_{b_1} \ket{\alpha_2}_{b_2} \to \ket{(\alpha_1 + \alpha_2)/\sqrt{2}}_{b_3} \ket{(\alpha_1-\alpha_2)/\sqrt{2}}_{b_4}$ to the pulses in modes $b_1$ and $b_2$;
(6) Bob applies a QND measurement to pulses $b_3$ and $b_4$ in order to execute a projective measurement $\{ \hat{Q}_s^{b_3 b_4},\hat{1}^{b_3 b_4}- \hat{Q}_s^{b_3 b_4} \}$ with $\hat{Q}_s^{b_3 b_4}:=\hat{1}^{b_3 b_4}-\sum_{n=0}^\infty \ket{n}\bra{n}_{b_3} \otimes \ket{n}\bra{n}_{b_4}$;
(7) If Bob receives an outcome corresponding to the projection $ \hat{Q}_s^{b_3 b_4}$, Bob declares the success of the protocol. 

In the virtual protocol for this scheme, since Bob's operations in steps (3)-(7) commute with the phase-flip channel $\Lambda^A_{|\braket{u_1}{u_0}|}$, the operations are assumed to be directly applied to the state $\ket{\psi'}_{Ab}$. In this sense,
the state after step (6) is described by $\ket{\chi}_{ABb_3b_4}=[\ket{00}_{AB}\ket{i \gamma_+}_{b_3} \ket{-i \gamma_-}_{b_4}+\ket{01}_{AB} \ket{-i \gamma_-}_{b_3} \ket{i \gamma_+}_{b_4} +\ket{10}_{AB} \ket{i \gamma_-}_{b_3} \ket{-i \gamma_+}_{b_4}+\ket{11}_{AB} \ket{-i \gamma_+}_{b_3} \ket{i \gamma_-}_{b_4}]/2$ with $\gamma_{\pm}:=[(\beta \pm \alpha) \sin(\theta/2) ]/\sqrt{2}$. This state can be represented, in the respective phase spaces of modes $b_3$ and $b_4$, by $\ket{\chi}_{ABb_3b_4}$ in Fig.~\ref{lo1} (a). This figure suggests an intuitive reason why this protocol can generate entanglement between qubits $AB$: 
If there are more photons in mode $b_3$ ($b_4$) than in mode $b_4$ ($b_3$), the possibility that the state has lived in the subspace spanned by $\{\ket{00}_{AB},\ket{11}_{AB} \}$ ($\{\ket{01}_{AB},\ket{10}_{AB} \}$) is higher.
A direct calculation shows $||{}_A \bra{j}  \hat{Q}_s^{b_3 b_4}  \ket{\chi}_{AB b_3 b_4}||^2= [1- e^{-\gamma_+^2 -\gamma_-^2} I_0(2 \gamma_+ \gamma_-) ]/2 $ for $j=0,1$ and ${}_{ABb_3 b_4} \bra{\chi} 
( \ket{1} \bra{0}_A \otimes  \hat{Q}_s^{b_3 b_4} ) \ket{\chi}_{ABb_3 b_4} = [e^{-(\gamma_+-\gamma_-)^2}- e^{-\gamma_+^2 -\gamma_-^2} I_0(2 \gamma_+ \gamma_-) ]/2 $, where 
$I_0(x):=\sum_{n=0}^\infty (x/2)^{2n} /(n!)^2$ is a modified Bessel function. 
Thus, the success probability $P_s$ is
\begin{equation}
P_s=1-e^{- (\beta^2+\alpha^2 ) \sin^2 (\theta/2) } I_0((\beta^2-\alpha^2) \sin^2 (\theta/2)). \label{eq:Ps}
\end{equation}
In addition, since the final state is written $\Lambda^A_{u_\alpha}(\ket{\phi}\bra{\phi}_{AB b_3 b_4} )$ with $\ket{\phi}_{AB b_3 b_4} :=  \hat{Q}_s^{b_3 b_4} \ket{\chi}_{ABb_3b_4}/ \sqrt{P_s} $ and $u_{\alpha}:=e^{-2\alpha^2 \sin^2 (\theta/2)}$,
it is concluded that
the concurrence $C$ between $A$ and $B b_3 b_4$ satisfies 
$C( \Lambda^A_{u_{\alpha}}( \ket{\phi}\bra{\phi}_{AB b_3 b_4} ))=C_{u_\alpha}^{\rm opt}(P_s)$
from Ref.~\cite{Appendix}.
On the other hand, for any $\alpha$ and $P$ satisfying $1-u_{\alpha} \le P < 1$, there is a choice of $\beta$ for making $P_s =P$ hold.
Hence, fixing $P_s =P$, we can choose $\alpha$ such that $u_\alpha$ is equivalent to $u^*$ of Eq.~(\ref{eq:uopt}).
Thus, the present protocol attains the optimal performance $C^{\rm opt}_{u^*}(P_s)$.

{\it Near-optimal protocol.}---We have shown that a protocol employing the QND measurement on incoming pulses can optimally generate entanglement between Alice's qubit $A$ and Bob's entire system $B b_3 b_4$ including pulses $b_3 b_4$.
However, in practice, it is difficult to
achieve such a QND measurement,
and the pulses $b_3 b_4$ are unsuitable for storing the entangled state for a long time.
Therefore, it is important to find a protocol that does not need to use a QND measurement and produces entanglement between Alice and Bob's qubits $AB$ instead of $A$ and $B b_3 b_4$.
One such protocol can be obtained by replacing steps (6) and (7) in the optimal protocol with the following steps [see Fig.~\ref{lo1} (a)]:
(6') Bob counts the number of photons by using photon-number-resolving detectors in modes $b_3$ and $b_4$, respectively;
(7') If the outcomes $m$ and $n$ of the two detectors are different,
Bob declares the success of the protocol.
We consider this modified protocol below.

From the definition, the success probability $P_s$ must be the same as Eq.~(\ref{eq:Ps}).
In the virtual protocol for this scheme, with probability 
$P_{mn}:= e^{-\gamma_+^2 -\gamma_-^2} (\gamma_+^{2m} \gamma_-^{2n}+\gamma_-^{2m} \gamma_+^{2n})/(2 m!n!)$,
the protocol returns outcomes $m$ and $n$, and provides a final state $\Lambda^A_{u_\alpha}( \ket{\phi_{mn}}\bra{\phi_{mn}}_{A B})$ for state $\ket{\phi_{mn}}_{AB}:=  {}_{b_3} \bra{m} _{b_4} \bra{n} \ket{\chi}_{AB b_3 b_4} /  \sqrt{P_{mn}}$
with concurrence 
$
C(\Lambda^A_{u_\alpha}( \ket{\phi_{mn}}\bra{\phi_{mn}}_{A B}))
=  u_\alpha^{\frac{1-T}{T}} e^{-\gamma_+^2-\gamma_-^2}|\gamma_+^{2m} \gamma_-^{2n}-\gamma_-^{2m} \gamma_+^{2n}|/(2 m! n! P_{mn})$
from Ref.~\cite{Appendix}.
Hence, for an entanglement monotone $E$ with $E(C)$,
the average of the entanglement monotones is determined by
$
\bar{E}=[\sum_{m,n\ge0} (1-\delta_{mn}) P_{mn}E(C(\Lambda^A_{u_\alpha}( \ket{\phi_{mn}}\bra{\phi_{mn}}_{A B})))]/P_s
$. Parameters $\alpha$ and $\beta$ (determining $\gamma_\pm$) should be chosen to maximize $\bar{E}$ with $P_s$ fixed.

In Fig.~\ref{lo1} (b), we show the performance of several known protocols \cite{L06,L08,M08,A09} as well as the optimal and near-optimal protocols in terms of the average concurrence $\bar{C}$. 
For comparison, we assume that all the devices used in the protocols are ideal.
From the figures, we can confirm that the near-optimal protocol performs similarly to the optimal protocol and it outperforms the existing protocols \cite{C06,L06,L08,M08,A09}.
Through the relation $E=E(C)$ for qubits, one could also easily estimate the performance even in terms of the entanglement monotone $E$.

In conclusion,
we have provided an optimal bound $E(C^{\rm opt}_{u^*}(P_s))$ defined by Eqs.~(\ref{eq:upper1}) and (\ref{eq:uopt})
for arbitrary entanglement manipulation via coherent-state transmission.
In addition, we have presented a simple optimal scheme and its practical version [Fig.~\ref{lo1} (a)] with almost optimal performance. 
This suggests that quantum optical devices in quantum communication can become as powerful as arbitrary operations.
The setting of the problem respects a shared nature of known realistic schemes \cite{B92,C06,L06,L08,M08,A09,ATKI10}, but we believe that our solution to the problem will provide new insights into fundamental theories such as those in Refs.~\cite{WPG07,JP99,W98,LP01} and into limits on other quantum communication protocols as in Refs.~\cite{V10,S11}.

We thank M.~Koashi, who pointed out the possibility of simplifying the proof of our proposition, W.~J.~Munro, M.~Owari, and K.~Tamaki, whose comments helped us to improve this paper, and K.~Igeta, N.~Matsuda, F.~Morikoshi, N.~Sota, and Y.~Tokura for helpful discussions. K.A. is supported by a MEXT Grant-in-Aid for Scientific Research on Innovative Areas 21102008.


\begin{thebibliography}{99}
\bibitem{cont}
Even if a system includes an observable with a continuous spectrum, the measurement outcomes from the system can be regarded as discrete 
by dividing the spectrum into intervals, and the following proof for an optimal bound is valid as long as we can take a proper entanglement monotone applicable to infinite dimensional systems.
\bibitem{examples}
The entanglement of formation, the geometric measure of entanglement, the Bures measure of entanglement, and the Groverian measure of entanglement are 
examples of such an entanglement monotone $E$ (cf.~\cite{SKB10}).
\bibitem{unital}
Random unitary channels on a qubit are equivalent \cite{GW03} to unital qubit channels.
\bibitem{converse}
In this proposition, Bob can be regarded as a sender who prepares qubits $AB$ in a state $\ket{\varphi}_{AB}$ and sends qubit $A$ to Alice through a unital qubit channel ${\cal E}^A$.
In this regard, since Bob's measurement merely corresponds to a modification of the prepared state $\ket{\varphi}_{AB}$,
this proposition implies that his state preparation is rather important.
\bibitem{Appendix}
If we input 
$
\ket{\omega}_{AB}= \sum_{j=0,1} \sqrt{p_j} \ket{j}_A \ket{\phi_j}_B
$ to a phase flip channel $\Lambda^A(\hat{\rho}):=f \hat{\rho} + (1-f) \hat{Z}^A \hat{\rho} \hat{Z}^A$ with $1/2 \le f \le 1$, 
the concurrence $C$ of the output state $\Lambda^A (\ket{\omega} \bra{\omega}_{AB})$ is
$
2(2f-1) \sqrt{p_0 p_1 (1-|\braket{\phi_1}{\phi_0}|^2)}
$
from the formula \cite{W98}.
\end{thebibliography}
\end{document}